\documentclass[showpacs,draft,floatfix]{revtex4}
\usepackage{graphicx}
\usepackage{latexsym}
\usepackage{amsmath}
\usepackage{amsfonts}
\usepackage{amssymb}
\begin{document}
\title{Normal and anti-normal ordered expressions for annihilation and creation
operators}
\author{J.M. Vargas-Mart\'{\i}nez and H. Moya-Cessa}
\affiliation{Instituto Nacional de Astrof\'{\i}sica, Optica y Electr\'onica, Apdo.
Postal 51 y 216, 72000 Puebla, Pue., Mexico}

\date{\today}
\begin{abstract}
We give the normal and anti-normal order expressions of the number
operator to the power $k$ by using the commutation relation
between the annihilation and creation operators. We use those
expressions to give  general formulae for  functions of the number
operator in normal and anti-normal order.
\end{abstract}

\maketitle

\section{Introduction}

In some problems in quantum mechanics it is needed to calculate
functions of the operator $\hat{n}= \hat{a}^{\dagger}\hat{a}$
where $\hat{a}$ and $\hat{a}^{\dagger}$ are annihilation and
creation operators of the harmonic oscillator, respectively. For
instance in ion traps \cite{wine} it is usual to have associated
Laguerre polynomials of order $\hat{n}$ \cite{Moya,Jonathan}.

Very recently Fujii and Suzuki have shown ordering expressions for
$\hat{n}^k$ as different types of polynomials with respect to the
number operator \cite{Suzuki}. They have shown nontrivial
relations including the use of Stirling numbers of the first kind
\cite{Abramowitz}.

Here we in fact do the opposite: we obtain an expression for
$\hat{n}^k$ in normal order (the antinormal order is then
straightforward, as it will be given in terms of similar
coefficients \cite{Vargas2}), i.e.  a sum of  coefficients
multiplying  normal ordered forms of $\hat{a}$ and
$\hat{a}^{\dagger}$. This allows us to obtain an expression for
the normal ordered form of a function of the operator $\hat{n}$
and demonstrate as a particular example a lemma in Louissel's book
for the exponential of the number operator \cite{Louissel}.

\section{Normal ordering}

One may use the commutation relations of the annihilation and
creation operators to obtain the powers of $\hat{n}$ in normal,
antinormal or symmetric order \cite{Louissel}. For instance, we
can express $\hat{n}^k$ in normal order, for $k=2$ as
\begin{equation}
\hat{n}^2 =  [\hat{a}^{\dagger}]^2\hat{a}^2 +
\hat{a}^{\dagger}\hat{a},
\end{equation}
for $k=3$ as
\begin{equation}
\hat{n}^3 =  [\hat{a}^{\dagger}]^3\hat{a}^3 +
3[\hat{a}^{\dagger}]^2\hat{a}^2 + \hat{a}^{\dagger}\hat{a},
\end{equation}
and for $k=4$
\begin{equation}
\hat{n}^4 =  [\hat{a}^{\dagger}]^4\hat{a}^4 + 6
[\hat{a}^{\dagger}]^3\hat{a}^3 + 7[\hat{a}^{\dagger}]^2\hat{a}^2
+\hat{a}^{\dagger}\hat{a},
\end{equation}
where the coefficients multiplying the different powers of the
normal ordered operators do not show an obvious form to be
determined. In writing the above equations we have  used
repeatedly the commutator $[\hat{a}, \hat{a}^{\dagger}]=1$. We may
infer that the coefficients in the above equations are Stirling
numbers of the second kind  (see also \cite{Solo}), i.e. we obtain
\begin{equation}
\hat{n}^k = \sum_{m=0}^k S_k^{(m)} [\hat{a}^{\dagger}]^m\hat{a}^m, \label{nalak}
\end{equation}
with \cite{Abramowitz}
\begin{equation}
S_k^{(m)}
=\frac{1}{m!}\sum_{j=0}^m(-1)^{m-j}\frac{m!}{j!(m-j)!}j^k.
\label{stirling}
\end{equation}
We now write a function of $\hat{n}$ in a Taylor series as
\begin{equation}
f(\hat{n}) = \sum_{k=0}^{\infty}\frac{f^{(k)}(0)}{k!}\hat{n}^k,
\end{equation}
and inserting (\ref{nalak}) in this equation we obtain
\begin{equation}
f(\hat{n}) = \sum_{k=0}^{\infty}\frac{f^{(k)}(0)}{k!}\sum_{m=0}^k
S_k^{(m)} [\hat{a}^{\dagger}]^m\hat{a}^m. \label{fofn}
\end{equation}
Because $S_k^{(m)}=0$ for $m > k$ we can take the second sum in
(\ref{fofn}) to infinite and interchange the sums to have
\begin{equation}
f(\hat{n}) = \sum_{m=0}^{\infty}[\hat{a}^{\dagger}]^m\hat{a}^m
\sum_{k=0}^{\infty}\frac{f^{(k)}(0)}{k!} S_k^{(m)} . \label{fofn2}
\end{equation}
For the same reason stated above, we may start the second sum at
$k=m$,
\begin{equation}
f(\hat{n}) = \sum_{m=0}^{\infty}[\hat{a}^{\dagger}]^m\hat{a}^m
\sum_{k=m}^{\infty}\frac{f^{(k)}(0)}{k!} S_k^{(m)}  .\label{fofn3}
\end{equation}
By noting that
\begin{equation}
\frac{\Delta^m f(x)}{m!} =
\sum_{k=m}^{\infty}\frac{f^{(k)}(x)}{k!}S_k^{(m)}, \label{Delta}
\end{equation}
where $\Delta$ is the difference operator, defined as
\cite{Abramowitz}
\begin{equation}
\Delta^m f(x) =  \sum_{k=0}^m (-1)^{m-k}\frac{m!}{k!(m-k)!}f(x+k),
\end{equation}
we may write (\ref{fofn3}) as
\begin{equation}
f(\hat{n}) =
\sum_{m=0}^{\infty}\frac{[\hat{a}^{\dagger}]^m\hat{a}^m
\Delta^m}{m!} f(0) \equiv :e^{\Delta \hat{n}}: f(0) \label{fofn4}
\end{equation}
where $:\hat{n}:$ stands for normal order.

\subsection{Lemma 1}

If we choose the function $f(\hat{n})=\exp(-\gamma \hat{n})$, we
have that
\begin{equation}
\Delta^m f(0) =
\sum_{k=0}^m(-1)^{m-k}\frac{m!}{k!(m-k)!}e^{-\gamma k},
\end{equation}
and then we obtain the well-known lemma \cite{Louissel}
\begin{equation}
e^{-\gamma \hat{n}} = :e^{(e^{-\gamma}-1)\hat{n}}:.
\end{equation}

\section{Anti-normal ordering}
Following the procedure introduced in the former section, we can
write $\hat{n}^k$ in anti-normal order as
\begin{equation}
\hat{n}^k = (-1)^k\sum_{m=0}^k (-1)^m S_{k+1}^{(m+1)}
\hat{a}^m[\hat{a}^{\dagger}]^m, \label{analak}
\end{equation}
and a function of the number operator as
\begin{equation}
f(\hat{n}) =
\sum_{m=0}^{\infty}(-1)^m\hat{a}^m[\hat{a}^{\dagger}]^m
\sum_{k=m}^{\infty}(-1)^k\frac{f^{(k)}(0)}{k!} S_{k+1}^{(m+1)}.
\label{afofn3}
\end{equation}
The second sum differs from (\ref{Delta}) in the extra $(-1)^k$
and the parameters of the Stirling numbers. We can define $u=-x$,
such that $f^{(k)}(x)_{x=0}= (-1)^k f^{(k)}(u)_{u=0}$, and use the
identity \cite{Abramowitz}
\begin{equation}
 S_{k+1}^{(m+1)} = (m+1)  S_{k}^{(m+1)} +  S_{k}^{(m)}
\end{equation}
to write
\begin{eqnarray}
f(\hat{n}) &=&
\sum_{m=0}^{\infty}(-1)^m\hat{a}^m[\hat{a}^{\dagger}]^m \\
\nonumber & &
\left((m+1)\sum_{k=m}^{\infty}\frac{f^{(k)}(u=0)}{k!}
S_{k}^{(m+1)} + \sum_{k=m}^{\infty}\frac{f^{(k)}(u=0)}{k!}
S_{k}^{(m)} \right)
\end{eqnarray}
so we can use again Eq. (\ref{Delta}) to finally write
\begin{equation}
f(\hat{n}) = (1+\Delta)\vdots e^{-\Delta \hat{n}}\vdots f(0)
\end{equation}
where $\vdots \hat{n} \vdots$ stands for anti-normal order.

\subsection{Lemma 2}

Let us consider again the function $f(\hat{n})=\exp(-\gamma
\hat{n})$. This gives us that $f(x)=e^{-\gamma x}$ and
$f(u)=e^{\gamma u}$. Therefore
\begin{equation}
\Delta^m f(u=0) =
\sum_{k=0}^m(-1)^{m-k}\frac{m!}{k!(m-k)!}e^{\gamma
k}=(e^{\gamma}-1)^m,
\end{equation}
such that we can obtain the exponential of the number operator in
anti-normal order (lemma) as
\begin{equation}
e^{-\gamma\hat{n}} = e^{\gamma} \vdots e^{(1-e^{\gamma})
\hat{n}}\vdots  . \label{lemma2}
\end{equation}
\subsubsection{Coherent states.}

Let us use Eq. (\ref{lemma2}) to find averages for coherent
states, $|\alpha\rangle = \hat{D}(\alpha)|0\rangle$, where
$\hat{D}(\alpha)=e^{\alpha\hat{a}^\dagger-\alpha^*\hat{a}}$ is the
so-called displacement operator and $|0\rangle$ is the vacuum
state:
\begin{equation}
\langle \alpha |e^{-\gamma\hat{n}}|\alpha \rangle = e^{\gamma}
\langle \alpha | \sum_{m=0}^{\infty}\frac
{(1-e^{\gamma})^m}{m!}\hat{a}^m [\hat{a}^{\dagger}]^m |\alpha
\rangle
\end{equation}
by using that
\begin{eqnarray}
\langle \alpha |\hat{a}^m [\hat{a}^{\dagger}]^m |\alpha \rangle
&=& \langle 0|(\hat{a}+\alpha)^m(\hat{a}^{\dagger}+\alpha^*)^m
|0\rangle \\ \nonumber &=& \sum_{k=0}^m
|\alpha|^{2k}\left(\frac{m!}{(m-k)!k!}\right)^2(m-k)!
\end{eqnarray}
we may write
\begin{equation}
\langle \alpha |e^{-\gamma\hat{n}}|\alpha \rangle = e^{\gamma}
\sum_{m=0}^{\infty}(1-e^{\gamma})^m L_m(-|\alpha|^2)
\end{equation}
where $L_m(x)$ are the Laguerre polynomials of order $m$. We can
finally write a closed expression for the sum above \cite{Grad} to
obtain the expected result for coherent states
\begin{equation}
\langle \alpha |e^{-\gamma\hat{n}}|\alpha \rangle =
e^{|\alpha|^2(e^{-\gamma}-1)}.
\end{equation}
\subsubsection{Fock states.}

For Fock or number states we obtain
\begin{eqnarray}
\langle n |e^{-\gamma\hat{n}} |n \rangle &=& e^{\gamma} \langle n
| \sum_{m=0}^{\infty}\frac{(1-e^{\gamma})^m}{m!}\hat{a}^m
[\hat{a}^{\dagger}]^m |n \rangle \\ \nonumber &=& e^{\gamma}
\sum_{m=0}^{\infty}\frac{(1-e^{\gamma})^m}{m!}\frac{(m+n)!}{n!}
\end{eqnarray}
rearranging the sum above with $k=n+m$ we have
\begin{equation}
\langle n |e^{-\gamma\hat{n}} |n \rangle = e^{\gamma}
\sum_{k=n}^{\infty}(1-e^{\gamma})^{k-n}\frac{k!}{n!(k-n)!}
\end{equation}
which has a closed expression, as
$\sum_{k=n}^{\infty}x^{k-n}\frac{k!}{n!(k-n)!}=(1-x)^{-n-1}$
\cite{Abramowitz}:
\begin{equation}
\langle n |e^{-\gamma\hat{n}} |n \rangle = e^{-\gamma n}
\end{equation}
\section{Conclusions}

In conclusion, we have written the normal and anti-normal order
expressions of $\hat{n}^k$ by using the commutation relation
between the annihilation and creation operators. The  coefficients
for such expressions  are the Stirling numbers of the second kind
\cite{Solo}. We then have used the difference operator to write a
function (that may be developed in Taylor series) of the number
operator in normal and anti-normal order, showing consistency with
the particular case of the exponential function lemma in normal
order.

This work was supported by Consejo Nacional de Ciencia y
Tecnolog\'{\i}a.

\end{document}